\documentclass[3p,times,procedia]{elsarticle}
\usepackage{nupha_ecrc}

\volume{00}

\firstpage{1}

\journalname{Nuclear Physics A}

\runauth{Ziyue Wang et al.}

\jid{nupha}

\jnltitlelogo{Nuclear Physics A}

\usepackage{amssymb}
\usepackage[figuresright]{rotating}

\begin{document}

\begin{frontmatter}



%% Instructions from Editor: Please use the following \dochead only in the preprint version (e-print arXiv etc.); 
%% use empty \dochead{} when submitting to Nuclear Physics A!
\dochead{XXVIIIth International Conference on Ultrarelativistic Nucleus-Nucleus Collisions\\ (Quark Matter 2019)}
%% e.g. \dochead{17th International Conference on Dynamical Processes in Excited States of Solids}

\title{Mass Correction to Chiral Kinetic Equations}


\author{Ziyue Wang$^1$, Xingyu Guo$^2$, Shuzhe Shi$^3$ and Pengfei Zhuang$^1$}

\address{$^1$Physics Department, Tsinghua University, Beijing 100084, China\\
             $^2$Institute of Quantum Matter, South China Normal University, Guangzhou 510006, China\\
             $^3$Department of Physics, McGill University, 3600 University Street, Montreal, QC, H3A 2T8, Canada}

\begin{abstract}
We study fermion mass correction to chiral kinetic equations in electromagnetic fields. Different from the chiral limit where fermion number density is the only independent distribution, the number and spin densities are coupled to each other for massive fermion systems. To the first order in $\hbar$, we derived the quantum correction to the classical on-shell condition and the Boltzmann-type transport equations. To the linear order in the fermion mass, the mass correction does not change the structure of the chiral kinetic equations and behaves like additional collision terms. While the mass correction exists already at classical level in general electromagnetic fields, it is only a first order quantum correction in the study of chiral magnetic effect.
\end{abstract}

\begin{keyword}
chiral kinetic equation\sep massive fermion \sep equal-time Wigner function 
\end{keyword}

\end{frontmatter}

% \linenumbers

\ \\

The Chiral Magnetic Effect~\cite{Fukushima:2008xe} has triggered a lot of interests in nuclear physics~\cite{Liao:2014ava} and condensed matter physics~\cite{Li:2014bha}. Three ingredients are crucial for the generation of the chiral magnetic effect, the magnetic field, the presence of chiral imbalance, and the massless fermions. In high energy heavy ion collisions which are expected to be a way of realizing the chiral magnetic effect, the coexistence of the first two ingredients may occur in the quark matter created in the initial stage of the collisions. For an out-of-equilibrium system in heavy ion collision, a natural way to describe the transport phenomena is through the kinetic theory in Wigner function formalism~\cite{Elze:1989un}. The chiral magnetic effect in out-of-equilibrium state in chiral limit is recently widely studied in the framework of kinetic theory, for instance~\cite{Hidaka:2018ekt}. By applying semiclassical expansion method to the kinetic equations, to the first order in $\hbar$, the chiral anomaly related effects are incorporated into the transport equation for the chiral fermion distribution function~\cite{Gao:2018wmr}. The transport equation is also applied to phenomenologically study of the charge separation in the pre-thermal stage of heavy ion collisions~\cite{Huang:2018wdl}. However, all quarks in QCD are massive, even in extremely hot quark matter. To check the degree of chiral anomaly in a real fermion system, it is necessary to study the fermion mass effect on the chiral magnetic effect. This is not a trivial problem even in the case of small fermion mass. With nonzero mass, fermions with different helicity are coupled to each other. It is of fundamental necessity to find out how finite mass modifies the chiral anomaly effects. There are already several attempts to study the kinetic equation of massive fermion~\cite{Weickgenannt:2019dks,Hattori:2019ahi,Wang:2019moi}.

To extract particle distribution functions in the electromagnetic background from the Wigner function and solve the kinetic equations as an initial value problem, one  introduces the equal-time Wigner function~\cite{BialynickiBirula:1991tx}
\begin{equation}
\label{w3}
{\cal W}(x,{\bf p})=\int d^3y e^{ipy}\left\langle\psi(x_+)e^{iQ\int_{-1/2}^{1/2}dsA(x+sy)y}\psi^\dag(x_-)\right\rangle.
\end{equation}
The equal-time Wigner function is not Lorentz covariant and is related to the covariant one through the energy integration, ${\cal W}(x,{\bf p}) =\int dp_0 W(x,p)\gamma^0$. Since the equal-time Wigner functions is not real, the physical phase-space densities are defined through their spin components,
\begin{eqnarray}
\label{decom}
{\cal W} &=& \frac{1}{4}\left[f_0+\gamma_5 f_1-i\gamma_0\gamma_5 f_2+\gamma_0 f_3+\gamma_5\gamma_0{\bf \gamma}\cdot{\bf g}_0+\gamma_0{\bf \gamma}\cdot{\bf g}_1-i{\bf \gamma}\cdot {\bf g}_2-\gamma_5{\bf \gamma}\cdot {\bf g}_3\right].
\end{eqnarray}
By calculating the physical densities of the system like charge, energy, momentum and angular momentum in terms of the equal-time Wigner function, one can establish the physical meaning of the equal-time components~\cite{BialynickiBirula:1991tx}. For instance, $f_0$ is the charge density, $f_3$ the mass density, ${\bf g}_0$ the spin current, ${\bf g}_1$ the number current, and ${\bf g}_3$ the intrinsic magnetic moment. The equal-time transport equations of the components can be viewed as extension of the classical Boltzmann equation~\cite{Zhuang:1995pd,Guo:2017dzf},
\begin{eqnarray}
\label{transport}
\hbar(D_t f_0+{\bf D}\cdot{\bf g}_1) &=& 0,\qquad\qquad~ \hbar(D_t{\bf g}_0+{\bf D} f_1)-2{\bf \Pi}\times{\bf g}_1 ~=~ 0,\nonumber\\
\hbar(D_t f_1+{\bf D}\cdot{\bf g}_0) &=& -2m f_2,\qquad \hbar(D_t{\bf g}_1+{\bf D} f_0)-2{\bf \Pi}\times{\bf g}_0 ~=~ -2m{\bf g}_2,\nonumber\\
\hbar D_t f_2-2{\bf \Pi}\cdot{\bf g}_3 &=&2m f_1,\qquad~~~ \hbar(D_t{\bf g}_2-{\bf D}\times{\bf g}_3)+2{\bf \Pi} f_3 ~=~2m{\bf g}_1,\nonumber\\
\hbar D_t f_3-2{\bf \Pi}\cdot{\bf g}_2 &=& 0,\qquad\qquad~ \hbar(D_t{\bf g}_3+{\bf D}\times{\bf g}_2)+2{\bf \Pi} f_2 ~=~ 0,
\end{eqnarray}
and the equal-time constraint equations which are the extension of the classical on-shell condition~\cite{Zhuang:1995pd,Guo:2017dzf},
\begin{eqnarray}
\label{constraint}
&&\int dp_0 p_0 V_0-{\bf \Pi}\cdot{\bf g}_1+\Pi_0 f_0 = mf_3,\qquad\int dp_0 p_0 {\bf A}+\frac{1}{2}\hbar{\bf D}\times{\bf g}_1+{\bf \Pi}f_1-\Pi_0{\bf g}_0 = -m{\bf g}_3,\nonumber\\
&&\int dp_0 p_0 A_0+{\bf \Pi}\cdot{\bf g}_0-\Pi_0f_1 = 0,\qquad~~~ \int dp_0 p_0 {\bf V}-\frac{1}{2}\hbar{\bf D}\times{\bf g}_0+{\bf \Pi}f_0-\Pi_0{\bf g}_1 = 0,\nonumber\\
&&\int dp_0 p_0 P+\frac{1}{2}\hbar{\bf D}\cdot{\bf g}_3+\Pi_0f_2 = 0,\qquad \int dp_0 p_0 S^{0i}{\bf e}_i-\frac{1}{2}\hbar{\bf D}f_3+{\bf \Pi}\times{\bf g}_3-\Pi_0{\bf g}_2 = 0,\nonumber\\
&&\int dp_0 p_0 F-\frac{1}{2}\hbar{\bf D}\cdot{\bf g}_2+\Pi_0f_3 = mf_0,\quad \int dp_0 p_0 S_{jk}\epsilon^{jki}{\bf e}_i-\hbar {\bf D}f_2+2{\bf \Pi}\times{\bf g}_2+2\Pi_0{\bf g}_3 = 2m{\bf g}_0,
\end{eqnarray}
where the equal-time operators are defined as $D_t=\partial_t+Q\int_{-1/2}^{1/2}ds {\bf E}({\bf x}+is\hbar{\bf \nabla}_p)\cdot{\bf \nabla}_p$, ${\bf D}={\bf \nabla}+Q\int_{-1/2}^{1/2}ds{\bf B}({\bf x}+is\hbar{\bf \nabla}_p)\times{\bf \nabla}_p$, $\Pi_0 = iQ\hbar\int_{-1/2}^{1/2}dss{\bf E}({\bf x}+is\hbar{\bf \nabla}_p)\cdot{\bf \nabla}_p$ and ${\bf \Pi}={\bf p} -iQ\hbar\int_{-1/2}^{1/2}dss{\bf B}({\bf x}+is\hbar{\bf \nabla}_p)\times{\bf \nabla}_p$. In the equal-time framework, the electromagnetic field strengths ${\bf E}$ and ${\bf B}$ are used instead of the fields $A_\mu$. It is clear that, the constraint equations couple the equal-time components $f_i(x,{\bf p})$ and ${\bf g}_i(x,{\bf p})\ (i=0,1,2,3)$ with the first order energy moments $\int dp_0 p_0 \Gamma_a(x,p)\gamma_0$ with $\Gamma_a=\{F, P, V_\mu, A_\mu, S_{\mu\nu}\}$. Only in the classical limit with on-shell condition $p_0=\pm E_p$, the first order moments are reduced to $\pm E_p \{f_i,{\bf g}_i\}$, and the transport and constraint equations become a group of closed kinetic equations for the equal-time Wigner function. In general case with quantum off-shell effect, all the energy moments are independent, they couple to each other and form a hierarchy of kinetic equations~\cite{Zhuang:1998bqx}. To see explicitly the classical limit and quantum correction order by order, we make semiclassical expansions for the covariant and equal-time components and operators. 

Taking the classical on-shell condition, the constraint equations (\ref{constraint}) automatically determine the position of the shell, namely the particle energy $E_p=\sqrt{m^2+{\bf p}^2}$ and reduce the number of independent spin components. At the first quantum level, to include a general off-shell effect in the kinetic theory, we add a continuous function of $p_0$ to the classical on-shell condition. At both classical and first quantum level, only the fermion number density $f_0$ and spin current ${\bf g}_0$ are independent, and the other components can simply be expressed in terms of them~\cite{Zhuang:1995pd,Guo:2017dzf}. 

Taking the expansion of $\hbar$ accordingly and use the relations from the constraint equations, one can derive the transport equations of these two independent components at classical and quantum level, 
\begin{eqnarray}
\label{transportf0}
\left(D_t^{(0)} \pm {{\bf p}\over E_p}\cdot{\bf D}^{(0)}\right)f^{(0)\pm}_0 &=& 0,\nonumber\\
\left(D_t^{(0)} \pm {{\bf p}\over E_p}\cdot{\bf D}^{(0)}\right){\bf g}^{(0)\pm}_0 &=& {1 \over E_p^2}\left[{\bf p}\times \left({\bf E}\times {\bf g}^{(0)\pm}_0\right)\mp E_p{\bf B}\times{\bf g}^{(0)\pm}_0\right],\nonumber\\
\left(D_t^{(0)}\pm\frac{{\bf p}}{E_p} \cdot {\bf D}^{(0)}\right) f^{(1)\pm}_0 &=& \frac{{\bf E}}{2E_p^2}\cdot{\bf D}^{(0)}\times{\bf g}_0^{(0)\pm}\mp\frac{1}{2E_p^3}{\bf B}\cdot({\bf p}\cdot{\bf D}^{(0)}){\bf g}_0^{(0)\pm}+\frac{{\bf B}\times{\bf p}}{E_p^4}\cdot {\bf E}\times{\bf g}_0^{(0)\pm},\nonumber\\
\left(D_t^{(0)}\pm\frac{{\bf p}}{E_p}\cdot{\bf D}^{(0)}\right){\bf g}^{(1)\pm}_0 &=& {1 \over E_p^2}\left[{\bf p}\times \left({\bf E}\times {\bf g}^{(1)\pm}_0\right)\mp E_p{\bf B}\times{\bf g}^{(1)\pm}_0\right]\mp \left(\frac{\bf B}{2E_p^3}\pm\frac{{\bf E}\times{\bf p}}{2E_p^4}\right){\bf p}\cdot{\bf D}^{(0)} f^{(0)\pm}_0\nonumber\\
&&\mp \left(\frac{({\bf p}\cdot{\bf E})({\bf E}\times{\bf p})}{E^5_p}\pm\frac{{\bf p}\times({\bf B}\times{\bf E})}{2E_p^4}\right)f^{(0)\pm}_0.
\end{eqnarray}
The above four equations describe the transport of number density and spin density of a massive fermion system in the external electromagnetic field background, and are valid for arbitrary fermion mass. With an appropriate initial condition, one can solve firstly the classical transport equations and then the quantum transport equations order by order. The first two equations are the phase-space version of a generalized Bargmann-Michel-Telegdi equation~\cite{Bargmann:1959gz}. The particle number density and spin density are independent to each other at classical level but are coupled at quantum levels. 

In chiral limit, while the vector and axial vector currents $V_\mu$ and $A_\mu$ are coupled to each other, their combinations $J_\mu=V_\mu+A_\mu$ and $J_\mu=V_\mu-A_\mu$ are decoupled. The physics behind is the number conservation of left-handed and right-handed fermions. To see the mass correction to the chiral conservation, we still introduce the chiral currents $J_\mu^\chi=V_\mu+\chi A_\mu\ (\chi=\pm)$ in covariant formalism or $f_\chi=f_0+\chi f_1$ and ${\bf g}_\chi={\bf g}_1+\chi{\bf g}_0$ in equal-time formalism. For massless fermions with certain chirality, the spin is not an independent degree of freedom, and the spin distribution can be determined by the number density. For massive fermions, as the spin direction does not follow the momentum direction, ${\bf g}_\chi$ and $f_\chi$ independent components. 

Using the transport equations (\ref{transportf0}), we can derive the transport equation of the classical and quantum chiral components $f_\chi^{(0)\pm}$ and $\tilde f_\chi^{(1)\pm}=f_\chi^{(1)\pm}\mp\chi{{\bf p}\cdot{\bf B}\over 2E_p^3}f_\chi^{(0)\pm}$, where we have shifted the first-order distribution from $f_\chi^{(1)}$ to $\tilde f_\chi^{(1)}$ to remove the infrared divergence in chiral limit~\cite{Chen:2014cla}. To see clearly the mass correction to the chiral kinetic equations, we have taken Taylor expansion in terms of the fermion mass $m$ in the transport equations and kept only the linear terms in $m$ which are explicitly shown on the right-hand side. The sum of the two transport equation of $f_\chi^{(1)}$ to $\tilde f_\chi^{(1)}$ leads to the transport equation for the chiral component $f_\chi=f_\chi^{(0)}+\hbar\tilde f_\chi^{(1)}$. Introducing berry curvature~\cite{Berry:1984jv} ${\bf b}=\chi{\bf p}/2p^3$, dispersion relation $\epsilon_p=p(1-\hbar Q{\bf B}\cdot{\bf b})$ and velocity ${\bf v}_p={\bf \nabla}_p\epsilon_p={\bf p}/p(1+2\hbar Q{\bf b}\cdot{\bf B})-\hbar Q({\bf p}/p\cdot{\bf b}){\bf B}$, the transport equation can be simplified as
\begin{equation}
\label{cktm}
\partial_tf_\chi^\pm+\dot{\bf x}\cdot{\bf \nabla} f_\chi^\pm+\dot{\bf p}\cdot{\bf \nabla}_p f_\chi^\pm = \chi m{F_1[{\bf g}_3^\pm]\over \sqrt G}+\hbar m\frac{F_2[{\bf g}_3^{(0)\pm}]}{\sqrt{G}},
\end{equation}
with the phase-space factor $G=(1+\hbar Q{\bf B}\cdot{\bf b})^2$ and the equations of motion $\dot{\bf x} = \frac{1}{\sqrt G}[{\bf v}_p+\hbar Q({\bf v}_p\cdot{\bf b}){\bf B}+\hbar Q{\bf E}\times{\bf b}]$, $\dot{\bf p} = \frac{Q}{\sqrt G}[{\bf v}_p\times{\bf B}+{\bf E}+\hbar({\bf E}\cdot{\bf B}){\bf b}]$. And the two functions of the magnetic moment ${\bf g}_3$ are defined as $F_1[{\bf g}_3] = -\frac{{\bf E}\cdot{\bf g}_3^\pm}{p^2}$ and $F_2[{\bf g}_3] = \pm\frac{1}{2p^3}{\bf D}^{(0)}\cdot({\bf E}\times{\bf g}_3^\pm)+\frac{1}{2p^4}({\bf p}\cdot{\bf D}^{(0)})({\bf B}\cdot{\bf g}_3^\pm)\mp\frac{3}{2p^5}({\bf p}\times{\bf B})\cdot({\bf E}\times{\bf g}_3^\pm)$. In comparison with the chiral kinetic equation for massless fermions~\cite{Son:2012zy}. The two kinetic equations with and without fermion mass have the same structure: the berry curvature, the equations of motion, and the phase-space factor are exactly the same. The only difference is the nonzero collision terms on the right-hand side generated by the interaction between the massive particle spin and electromagnetic fields.

Given the above kinetic equation (\ref{cktm}) for fermion systems with small mass, it is of great interest to find possible analytic solutions. When we turn off the electric field and keep only the magnetic field, corresponding to the physics of chiral magnetic effect, the effective collision term can be solved through the classical transport equation. In this case, the effective collision term $\beta({\bf x},{\bf p},t)\equiv\hbar mF_2[{\bf g}_3^{(0)}]/\sqrt G$ is known and the equation can be analytically solved~\cite{Yan:2006ve}. The solution can be in general written as $f_\chi^\pm({\bf x},{\bf p},t) = f_{\chi 0}^\pm({\bf x}_0({\bf x},{\bf p},t;t_0),{\bf p}_0({\bf x},{\bf p},t;t_0),t_0)+\int_{t_0}^t\beta({\bf x}({\bf x}_0,{\bf p}_0,t_0;t'),{\bf p}({\bf x}_0,{\bf p}_0,t_0,t'),t')dt'$. The other point in the case with only magnetic field is that, the mass correction is only a quantum correction, since the collision term is at the first order in $\hbar$. This leads to the conclusion that, the mass correction to the chiral magnetic effect should be small. When the electric field is turned on, the mass correction appears already at classical level, see the first collision term with $F_1$ in (\ref{cktm}). Therefore, in the case with only electrical field or both electrical and magnetic fields, the mass correction will become more important.

While the quantum chiral anomaly and related phenomena in fermion systems are widely discussed in chiral limit, the mass correction in real case should be seriously considered. We derived the transport equations for the particle number and spin densities at classical level and to the first order quantum correction. To see clearly the fermion mass correction to the chiral kinetic equations, we take Taylor expansion in terms of the mass, and to the linear order we obtained kinetic equations with mass correction. The mass correction is reflected as effective collision terms in the transport equations. Different from chiral limit where the chiral number density is the only independent quantity and its transport equation controls the evolution of the system, spin density becomes independent for massive fermions, and the chiral number density and spin density are coupled to each other. In the case with only magnetic field, the mass correction is a quantum correction, and the chiral number density can be analytically solved.



%% References


\bibliographystyle{elsarticle-num}
\bibliography{refs2.bib}






% \end{thebibliography}

\end{document}